\documentclass[aps,prb,groupedaddress,showpacs,twocolumn,letterpaper]{revtex4}
\usepackage{graphics}
\usepackage{epsfig}
\usepackage{graphicx}
\usepackage{dcolumn}

\begin{document}

\title{Crossover from diffusive to ballistic transport in
  semiconductor nanostructures} 
\author{Dan Csontos}
\email{csontos@phy.ohiou.edu}
\affiliation{Department of Physics and Astronomy, and Nanoscale and
  Quantum Phenomena
Institute, Ohio University, Athens, Ohio 45701-2979}
\author{Sergio E. Ulloa}
\affiliation{Department of Physics and Astronomy, and Nanoscale and
  Quantum Phenomena
Institute, Ohio University, Athens, Ohio 45701-2979}

\begin{abstract}
{\footnotesize We present a detailed microscopic study of
  quasi-ballistic transport in deep submicron semiconductor
  channels. In particular, we study the crossover between the
  diffusive and ballistic regimes of transport and identify
  signatures in the electrostatic response, electron density, and
  nonequilibrium electron distribution function that are due to
  ballistic and diffusive transport, respectively. Our
  theoretical and computational approach is based on the Boltzmann
  transport equation for a nondegenerate electron system, together
  with the Poisson  
  equation, from which we obtain the nonequilibrium electron
  distribution in a self-consistent way. We find that the electron
  distribution is significantly different from a near-equilibrium, shifted
  Maxwell-Boltzmann distribution function, and that it displays a
  large broadening, as 
  well as pronounced features, peaks and shoulders in the
  high-velocity tail of the distribution. These features are
  signatures of the nonequilibrium and quasi-ballistic nature of the
  electron transport. An analysis of the spatial density distributions
  of electrons in the semiconductor channel region shows that the
  relative scattering efficiency varies within the channel region. Our
  study of the crossover between the diffusive and ballistic regimes of
  transport, which we investigate by downscaling the size of the
  semiconductor channel, shows that a crossover occurs around a
  channel length specific to the material parameters, and is clearly
  accompanied by a redistribution of the voltage drop between the
  contacts and the channel region. For these transition channel
  lengths, we also observe clear signatures of the diffusive-to-ballistic
  crossover in the spatial electron density distribution that manifest
  as exponential-like and linear spatial dependences, indicative of
  diffusive and ballistic transport, respectively.}
\end{abstract}

\maketitle

\section{Introduction}

Advances in fabrication and processing of semiconductor materials on
the nanoscale are presently enabling the realization of deep submicron
and nanoscale semiconductor structures and devices. At these 
ultrasmall length scales, finite size effects, hot-electrons,
ballistic transport and microscopic doping variations are important in
determining transport properties. In addition, these systems can be very
easily brought out-of-equilibrium, as very high fields are established
even for low applied bias voltages.

The physics of electron transport in ultrasmall semiconductor systems
is thus very rich as well as complex. Drift-diffusion approaches
as well as approximations based on a drifted-Maxwell-Boltzmann
distribution break down and an adequate treatment of the
nonequilibrium distribution function is needed. In particular, the
high-energy tail of the distribution can have complex features that
are determined by the scattering processes and the electrostatics
within the systems as shown experimentally in Raman spectroscopy
measurements of 
the electron distribution in submicron, inhomogeneous III-V
systems.\cite{experiments} These experiments have shown
far-from-equilibrium features 
such as strongly broadened velocity distributions and interesting
high-velocity tail structure.\cite{experiments}  

Theoretically, there have been a few recent works that have shown that
high-velocity structure in the distribution function can be attributed
to ballistic 
effects.\cite{barangerPRB1984, barangerPRB1987, fischettiPRB1988,
  lundstromIEEE2002, rhewSSE2002, rahmanIEEE2003, 
  svizhenkoIEEE2003, sanoAPL2004, sanoPRL2004, csontosAPL2005} In a
recent paper by us,\cite{csontosAPL2005} we discussed the spatial
dependence of the distribution function in deep submicron
GaAs channels and showed how quasi-ballistic signatures manifest as a
broadening and high-velocity tail structure, as seen in the
experiments. We also analyzed microscopic features of scattering in
the channel region in the presence of energy-dependent scattering due
to polar-optical phonons. Interesting similar analyses with or without
the inclusion of finite scattering have also been previously reported
within the same theoretical frameworks.\cite{sanoAPL2004, sanoPRL2004,
  rhewSSE2002}   

The scattering problem in submicron and nanoscale channels is
complicated since scattering and the electrostatic response of
the system constitute a strongly coupled physical problem that creates
a self-consistent feedback and an interesting charge
redistribution. Since pure
ballistic transport is not attainable at room temperature in the
presence of finite scattering\cite{sanoAPL2004, sanoPRL2004} and thus,
the transport in 
submicron systems is quasi-ballistic to a varying degree, it is
interesting to ask: 
What are the signatures of nonequilibrium and quasi-ballistic
transport in quantities other than the electron distribution, such as
electron density, potential energy and how can one separate the
ballistic contribution to the problem and study its influence on the
total quantities? More importantly, what constitutes the crossover
between diffusive and quasi-ballistic transport and how does it
manifest on a microscopic level?

In this paper we report on a detailed study of the crossover between
the diffusive and quasi-ballistic regimes of transport in
(sub)micrometer scale semiconductor channels, in particular
investigating how this crossover manifests in the electrostatic
response of the system, the electron redistribution within the channel
region, the contribution of ballistic electrons to the total electron
density, and the nonequilibrium electron distribution function. Our
theoretical and computational framework is based on a self-consistent
solution of the Boltzmann and Poisson equations,\cite{csontosJCE2004}
which enables us to calculate nonequilibrium electron distributions
and corresponding moments, as well as the accurate electrostatic
reponse.    

We identify that, for the GaAs system that we study, with a given
doping profile and at room-temperature, a crossover to quasi-ballistic
transport occurs for channel lengths smaller than $\approx 1.5$
$\mu$m. In the electrostatics, the crossover manifests as as
redistribution of the potential drop from the channel to the contact
regions, for decreasing channel lengths. In the electron density, which
we study by separating components with different velocity, we observe
interesting spatial dependences reflecting the diffusive-to-ballistic
crossover. Finally, on a microscopic level, we examine the
nonequilibrium electron distribution function and its phase-space
dependence around the crossover regime. The presented results provide
insight into the role of scattering, ballistic transport and
electrostatics in the transport properties of deep submicron and
nanoscale semiconductor systems.

The paper is organized as follows: First, we give a brief introduction
to the theoretical model in Section \ref{model}, referring the reader
to Ref.\ \onlinecite{csontosJCE2004} for further details. Next, we will
present our results and discussion, first presenting an introduction
to the basic signatures of quasi-ballistic transport in Section
\ref{basics}. Subsequently, we discuss the crossover between the
diffusive and quasi-ballistic regimes of transport in Section
\ref{crossover}. We conclude and summarize our main results in Section
\ref{summary}. 

\section{Model}
\label{model}
A microscopic study of the kinetics of quasi-ballistic transport in
ultra-small inhomogeneous semiconductor structures requires in principle the
solution of the Boltzmann transport equation (BTE). In addition, in order to
capture the full physical picture, the BTE needs to be solved
self-consistently with the Poisson equation. This is a very difficult task
since the BTE is a nonlinear, integro-differential equation for the
semiclassical 
distribution function, $f(\mathbf{r},\mathbf{v},t)$, which in principle has
seven dimensions. Although the Monte Carlo method has been very
popular and successful for the solution of the BTE in semiconductor device
simulation,\cite{jacoboni, ravaioli} several
works\cite{csontosJCE2004, carrilloJCP2003, majoranaJCP2001} have 
recently solved the BTE by direct methods. This allows noise-free spatial
and temporal resolution of the electron distribution function, which in the
Monte Carlo method may be difficult to obtain due to the statistical nature
of the approach.

In our model, the electron transport is described by the following
one-dimensional BTE\cite{1DBTE} within the relaxation time approximation

\begin{equation}
-\frac{eE(x)}{m^{\ast }}\frac{\partial f(x,v)}{\partial v}+v\frac{\partial
f(x,v)}{\partial x}=-\frac{f(x,v)-f_{LE}(x,v)}{\tau },  \label{BTE}
\end{equation}%
where $m^{\ast }$ is the effective mass in the parabolic band approximation
and where we have assumed that the nonequilibrium electron distribution
relaxes to a local equilibrium distribution, $f_{LE}(x,v)$, which we in the
following assume to describe non-degenerate statistics according to the
following normalized Maxwell-Boltzmann (MB) distribution

\begin{equation}
f_{LE}(x,v)=n(x)\left[ \frac{m^{\ast }}{2\pi k_{B}T}\right]
^{1/2}\exp \left[ 
-\frac{m^{\ast }v^{2}}{2k_{B}T}\right] ,  \label{MB}
\end{equation}%
where $n(x)$ is the local electron density, and $T$ is the lattice
temperature. The electron density is obtained from the full electron
distribution according to

\begin{equation}
n(x)=\int f(x,v)dv.  \label{density}
\end{equation}%
The BTE, eq. (\ref{BTE}), is coupled to the Poisson equation through the
electric field and electron density. In the following we assume that our
system is unipolar and hence, that the Poisson equation is given by

\begin{equation}
\frac{d^{2}\phi }{dx^{2}}=-\frac{dE}{dx}=-e\frac{N_{D}(x)-n(x)}{\epsilon
\epsilon _{0}}~,  \label{poisson}
\end{equation}
where $N_{D}(x)$ is a given inhomogeneous doping distribution
profile and $\epsilon$ is the dielectric constant. 

Equations (\ref{BTE}-\ref{poisson}) are coupled through the electron density
and electric field, and thus, they need to be solved self-consistently. We
use a numerical approach based on finite difference and relaxation methods,
the details of which can be found in Ref.\
\onlinecite{csontosJCE2004}. As boundary 
conditions, we adopt the following scheme: For the potential, the values at
the system boundaries, $x_{l(eft)}$ and $x_{r(ight)}$, are fixed to
$\phi (x_{l})=V_{b}$ and $\phi (x_{r})=0$% 
, respectively, corresponding to an externally applied voltage $V_{b}$. The
electron density is allowed to fluctuate around the system boundaries
subject to the condition of global charge neutrality, which is enforced
between each successive iteration in the self-consistent Poisson-Boltzmann
loop. The size of the highly-doped contacts is chosen to be large enough
such that the electron density is constant deep inside the contacts. In
addition, the size of the contacts has to be large enough, such that the
electric field deep in the contacts is constant, and very low. This allows
us to use the analytical, linear response solution to the BTE

\begin{equation}
f(x_{l,r},v)=f_{LE}(x_{l,r},v)\left[ 1-vE(x_{l,r})\tau /k_{B}T\right] ,
\label{linBTE}
\end{equation}%
as phase space boundary conditions at $x_{l,r}$, where we use the local
equilibrium distribution, $f_{LE}(x_{l,r},v)$ and local electric field, $%
E(x_{l,r})$, obtained from the previous numerical solution to the
Poisson-Boltzmann iterative loop. At the velocity cut-off in phase space, we
choose $f(x,v_{\max })=f(x,-v_{\max })=f_{LE}(x,v)$, which is reasonable
since, in the calculations, we assume $v_{\max }\geq 30k_{B}T$. A more
detailed description and discussion of our numerical method can be found in
Ref.\ \onlinecite{csontosJCE2004}.

\section{Results and discussion}
\label{results}
In the following, we will present and discuss our main results. First,
we will discuss some basic signatures of quasi-ballistic transport,
followed by a study of the crossover between the diffusive and
quasi-ballistic regimes of transport. Our model
structure consists of a lightly doped region (hereafter called the channel),
with doping concentration $N=10^{19}$ m$^{-3}$, sandwiched between two
highly doped semiconductor contact regions with doping concentration $%
N^{+}=10^{23}$ m$^{-3}$. We assume that the entire $N^{+}-N-N^{+}$
structure consists of GaAs, and use the appropriate parameters in
the calculations (effective mass $m^{\ast }=0.067m_{0}$, where
$m_{0}$ is the bare electron mass, and dielectric constant
$\epsilon =13.1$). Unless otherwise indicated, we assume that the
lightly doped channel region is 200 nm long. In addition, all
of our calculations are performed at $T=300$ K, using a scattering time of $%
\tau =0.25$ ps, which corresponds to a realistic scattering time due to
polar optical phonon scattering.\cite{scattering} The microscopic
effects of an energy-dependent scattering mechanism have been
previously studied in Ref.\ \onlinecite{csontosAPL2005}.

\subsection{Basic signatures of quasi-ballistic transport}
\label{basics}
The description of transport in submicron, inhomogeneous semiconductor
structures such as the
ones described in our paper is a complicated process due to \textit{i)} the
size of the system, \textit{ii)} the inhomogeneous field distribution,
\textit{iii)} the strongly non-equilibrium and quasi-ballistic nature
of the problem,\textit{\ iv)} the
self-consistent feedback between the scattering processes and
electrostatics. In the following, we will discuss some general properties of
our model $N^{+}-N-N^{+}$ structure, which are valid for a wide range of
ultrasmall, inhomogeneous semiconductor structures and devices. Some of
these results have been previously discussed by us in Ref.
\onlinecite{csontosAPL2005}.

\begin{figure}[t]
\par
\begin{center}
\scalebox{0.5}{\epsfig{file=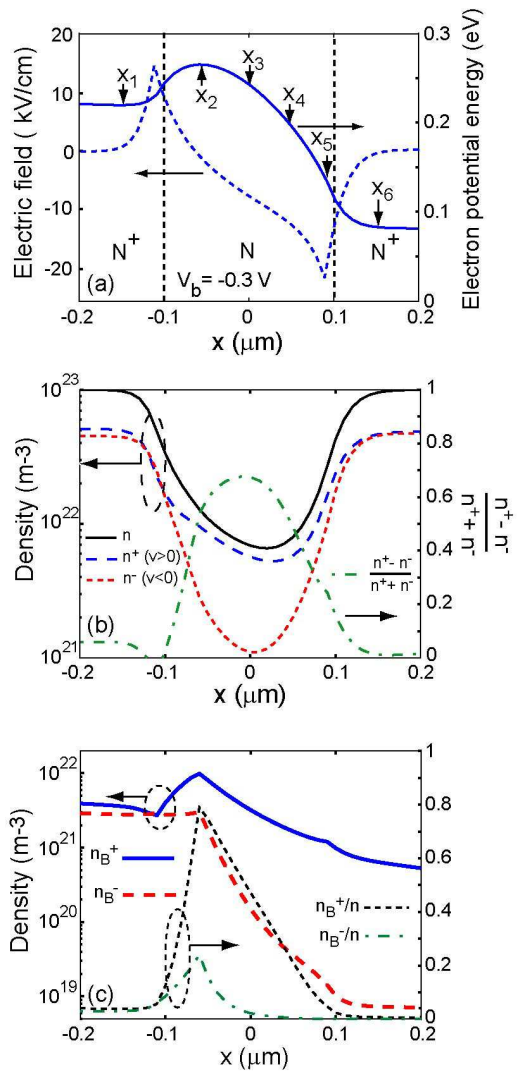}} \\
\end{center}
\caption{(Color online) (a) Electron potential energy and electric
field for a $N^{+}-N-N^{+}$ structure with doping
$10^{23}/10^{19}/10^{23}$ m$% 
^{-3}$ for an applied bias voltage of $V_{b}=-0.3$ V, temperature
$T=300$ K, and with scattering time $\tau =2.5\cdot 10^{-13}$ s.
Vertical dashed lines mark the interfaces between the highly doped
$N^{+}$ and lightly doped $N$ regions. (b) Electron
densities, $n^{+}$, $n^{-}$, $n$, and $(n^{+}-n^{-})/(n^{+}+n^{-})$ as
defined in the text. (c) ``Ballistic'' electron densities,
$n_{B}^{+}$, $n_{B}^{-}$, and ``ballistic'' electron fractions,
$n_{B}^{+}/n$, and $n_{B}^{-}/n$ as defined in the text.} \label{figure1}
\end{figure}

In Fig.\ \ref{figure1}(a) the electron potential energy (solid line) and
electric field (dotted line) are shown for a $N^{+}-N-N^{+}$ structure with
dimensions 2.9/0.2/2.9 $\mu $m, for an applied bias voltage of $V_{b}=-0.3$
V. The potential energy has a very slow spatial dependence in the large
contacts resulting in only $\approx 50\%$ of the voltage drop occuring
within
the contact regions. In contrast, most of the potential drops over the
lightly doped, deep submicron channel region, resulting in a dramatic spatial
variation. Correspondingly, a large field build-up, with opposite sign,
occurs at the $N^{+}/N$ interfaces, a strongly inhomogeneous electric field
is observed in the channel region, and in contrast, the field in the
contacts is very small and constant, thus validating the assumption of a
near-equilibrium solution to the BTE\ at the system boundaries. The
observed strong spatial dependence of the electron potential energy
and
electric field in the channel region is due to the carrier diffusion
induced by the large variation in doping concentration.\cite{csontosAPL2005}

In addition to the electrostatic picture discussed above, valuable
information can be obtained from the carrier distribution within the
system for a given applied external field and inhomogeneous doping profile.
In this context, it is interesting to study the separate contributions, $%
n^{+}$ due to electrons with $v\geq 0$ (hereafter denoted as $v^{+}$), and $%
n^{-}$ due to electrons with $v<0$ (hereafter denoted as $v^{-}$), to the
total electron density%
\begin{equation}
n(x)=n^{+}(x)+n^{-}(x)=\int_{v^{+}}f(x,v)dv+\int_{v^{-}}f(x,v)dv.
\label{totaldens}
\end{equation}%
In addition we also introduce and study the ``ballistic'' electron density,%
\begin{equation}
n_{B}(x)=\int_{\left\vert v\right\vert >v_{p}}f(x,v)dv\text{ ,}
\label{ballistic density}
\end{equation}%
which corresponds to the density of electrons with energies higher than the
top of the potential barrier between the source contact and the channel
[occurring at $x_{2}$ in Fig.\ \ref{figure1}(a)], thus corresponding to a
``ballistic'' electron density. In eq. (\ref{ballistic density}),
$v_{p}=\sqrt{2/m^{\ast }[U_{0}-U(x)]}$ and $U_{0}$ and $U(x)$ are the
barrier top and local electron potential energies, obtained from the
Poisson equation (\ref{poisson}), respectively. 

In Fig.\ \ref{figure1}(b) the electron densities $n^{+}$ (dashed line),
$n^{-}$ (dotted line), and $n=n^{+}+n^{-}$ (solid line), i.e., the
total electron density, are
shown for the structure with the same parameters as discussed in
Fig.\ \ref{figure1}(a). In addition, we show the spatial dependence of
the surplus of electrons with velocities $v^{+}$ relative to the total
electron density, $P_{n}=(n^{+}-n^{-})/(n^{+}+n^{-})$ (dashed-dotted
line), which can be viewed as a ``polarization'' of the electrons in
velocity space.

Several observations are made: First, the total electron
density, $n$, displays an asymmetric spatial dependence with respect to the
direction of transport, indicating a pile-up of charge at the source-channel
interface. Second, it is seen that the difference between the two
densities, $% 
n^{+}-n^{-}$, has a significant spatial dependence, and that the minimum of
the electron density components occurs at different points along the
channel. Furthermore, $P_{n}$ attains its maximum around mid-channel.

One can view the above results within the following intuitive picture:
As electrons are injected from the source contact into the channel,
they scatter at a rate given by $1/\tau $. However, the mean free path of
the $v^{+}$ and $v^{-}$ electrons is significantly different due to the
large built-in electric fields around the channel region, giving rise to a
considerable enhancement of the $n^{+}$ component of the density in
comparison to the $n^{-}$. The latter on the contrary is suppressed since,%
\textit{\ i)} electrons injected from the drain have to overcome a large
potential barrier, \textit{ii)} source-injected electrons
scattered from the $n^{+}$
contribution and back to the $n^{-}$ contribution of the density have a
relatively short mean free path due to
the built-in field that exerts a force in the opposite direction and thus,
quickly scatter back to the $n^{+}$ component. The spatial dependence
of $n^{+}$, $n^{-}$, and $(n^{+}-n^{-})/(n^{+}+n^{-})$ indicates that
the efficiency of the scattering is non-uniform
within the channel.

It is interesting to analyze the previously defined ``ballistic'' electron
density, $n_{B}$, and compare with the above shown results for the
electron densities calculated from the entire velocity range. In
Fig.\ \ref{figure1}(c) the thick solid and dashed lines
correspond to the $v^{+}$, and $v^{-}$ components (which we in the following
denote by $n_{B}^{+}$ and $n_{B}^{-}$) of $n_{B}$, respectively. In
addition, we also show the ``ballistic'' electron fractions
$n_{B}^{+}/n$ (thin dotted line), and $n_{B}^{-}/n$ (dashed-dotted line).

To the left
of the source-channel interface at $x=-0.1$ $\mu $m, the two
components, $n_{B}^{+}$ and $n_{B}^{-}$, have
a slow spatial dependence and are only slightly different in value,
reflecting the small field that is present in the source region. Around $%
x=x_{2}$, corresponding to the top of the potential barrier [see Fig.\ \ref
{figure1}(a)], the two components rapidly split, and $n_{B}^{+}$ peaks at the
barrier top, as expected by the pile-up of source-electrons at the
interface, where interestingly, only $\approx $80$\%$ of the total density
consists of electrons with positive velocity as seen in
$n_{B}^{+}/n$. Since the potential barrier
seen by the drain-injected electrons from the right is far too large
compared to $k_{B}T$
to explain the large $n_{B}^{-}$ component at $x_{2}$ in terms of
thermionically injected electrons
from the drain, the main contribution to that component consists of
backscattered source injected electrons. This is consistent with the
interesting analyses by Sano\cite{sanoAPL2004, sanoPRL2004} in which it
was shown mathematically that, due to the singular nature of the BTE (note,
that the analyses was made for the BTE in the relaxation time
approximation), a finite $n_{B}^{-}$ around the top of the barrier
separating the source and the channel region, is always present
provided there is a finite scattering rate. The intuitive physical
picture is that, 
electrons close to the conduction band edge can have an infinitely small
velocity, and thus can spend long enough time to experience a
scattering event. If
electrons incident from the source experience a sharp potential rise around
the barrier, electrons pile up close to the conduction band edge and are
successively slowed down, thus increasing the time spent around the barrier
region, and correspondingly giving rise to a significant contribution
in $n_{B}^{-}$. The contribution to $n^{-}$ becomes larger the larger
the curvature of the potential energy around the barrier
region.\cite{sanoPRL2004}

Beyond the location of the top of the barrier, for $x>x_{2}$,
$n_{B}^{-}$ drops rapidly within the channel region, followed by a
change in the slope past the channel-drain interface. The positive
component, $n_{B}^{+}$, on the other hand, has a much slower decay
within the channel region, followed by a similar change of decay rate
at the channel-drain interface as seen in $n_{B}^{-}$. We note that
$n_{B}^{+}$ is two orders of magnitude larger than $n_{B}^{-}$ at the
channel-drain interface showing that, indeed there are very few
drain-injected electrons at kinetic energies comparable to the
potential barrier at the source-channel interface.

A simple exponential fit to $n_{B}^{+}$ and $n_{B}^{-}$ in the channel
region shows that the mean free paths of electrons with $v>0$ and
$v<0$ are $\approx 140$ and 50 nm, respectively. Hence, source-injected
electrons have a mean free path comparable to the channel region and,
thus, propagate quasi-ballistically. However, inspite of the
quasi-ballistic transport of these electrons, their relative
contribution to the total electron density in the drain is very small, as
shown by the thin dashed and thin dashed-dotted line representing
$n_{B}^{+}/n$ and $n_{B}^{-}/n$ in
Fig.\ \ref{figure1}(c). As expected, the fractions $n_{B}^{+}/n$ and
$n_{B}^{-}/n$
reach their maximum at the barrier top and subsequently decrease
monotonically in the channel. We note, however, that
$n_{B}^{-}/n$ has an exponential-like decay rate, indicative of
diffusive-like transport, in contrast to $n_{B}^{+}/n$ which decays
almost linearly. 

\begin{figure}[t]
\par
\begin{center}
\scalebox{0.4}{\epsfig{file=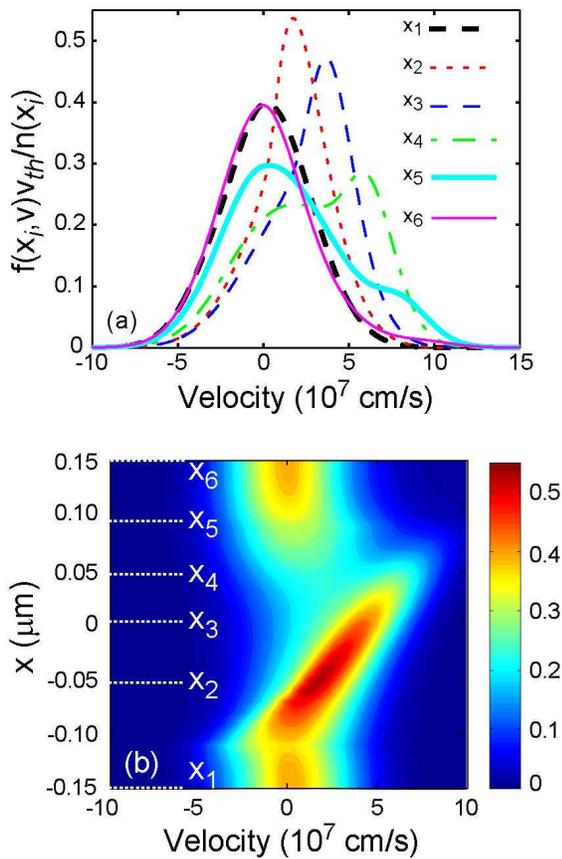}}
\end{center}
\caption{(Color online) (a) Normalized electron distribution,
$f(x_{i},v)v_{th}/n(x_{i})$, for the structure studied in Fig.\
\ref{figure1}, calculated at the spatial points $x_{1}$-$x_{6}$ as
marked by the arrows in Figs.\ \ref{figure1}(a) and
\ref{figure2}(b), for $V_{b}=-0.3$ V. (b) Contour plot of
$f(x,v)v_{th}/n(x)$ around the lightly doped $N$ region.} \label{figure2} 
\end{figure}

Full information about the microscopic details of the electron
transport is contained in the nonequilibrium electron distribution
function, to which we have access through our self-consistent solution
of 
the BTE. In Fig.\ \ref{figure2}(a) we show the normalized electron
distribution 
function, $f(x_{i},v)v_{th}/n(x_{i})$, where $v_{th}$ is the thermal
velocity and $n(x_{i})$ is the electron density at $x_{i}$, calculated at
different locations: one point in the source contact, four points along the
channel region, and one point in the drain contact [same as arrows in
Fig.\ \ref{figure1}(a)].
The electron distribution shows a strong spatial dependence, and a
far-from-equilibrium form at all but two of the spatial points depicted in
Fig.\ \ref{figure2}(a), namely the ones in the contacts. The thick dashed
curve, corresponding to the distribution
at $x_{1}=-0.15$ $\mu $m, is nearly Maxwellian, showing that the electrons
at that point are near equilibrium. This is in agreement with the fact that $%
x_{1}$ is situated in the contact region, where the field is very small and
constant [see Figs.\ \ref{figure1}(a) and \ref{figure2}(b)] and thus,
the electrons are distributed according to a shifted MB distribution,
as given by the linear response solution to the BTE, eq. (\ref{linBTE}).

The electron distributions at $x_{2}-x_{5}$ on the other hand are highly
asymmetric, displaying a strong weight toward positive velocities, as well
as pronounced structure in the high-velocity tail for $v^{+}$. At $
x_{2}=-0.05$ $\mu $m, the dotted curve shows the electron distribution at
the top of the potential barrier between the source and channel regions. The
distribution is distorted from a Gaussian-like MB distribution and shows a
suppression of electrons in the $v<0$ tail of the distribution as
expected since, intuitively, only electrons injected from the source,
with positive velocities, should be present. The presence
of a negative-velocity tail in the distribution at the top of the potential
barrier is again in accordance with our discussion above and the findings of
Sano,\cite{sanoAPL2004, sanoPRL2004} showing that significant
backscattering occurs around the barrier region due to the sharp
potential barrier and corresponding pile-up of
electrons at the conduction band edge resulting in an
increased backscattering efficiency due to the prolongued dwell time
in the barrier vicinity.

Beyond the barrier top, the peak of the
distribution rapidly shifts toward higher velocities and is significantly
broadened ($x_{3}=0$). In addition, beyond the channel half-point, the
low-velocity
contribution of the distribution function increases, and a second peak
emerges, as seen in the dash-dotted line, corresponding to $x_{4}=0.05$ $\mu
$m. At $x_{5}=0.09$ $\mu$m, the low-velocity peak dominates and only a
shoulder is observed on the high-velocity side of the non-equilibrium
distribution. Finally, at $x_{6}=0.15$ $\mu$m, i.e., in the drain
contact, the distribution function again assumes a near-equilibrium
form. Note however that a high-velocity tail is still observable, and
the distribution function, although closely resembling a shifted MB
distribution, is significantly different from the one at $x_{1}$.

A contour plot of $f(x,v)v_{th}/n(x)$ around the channel region is shown
in Fig.\ \ref{figure2}(b) and further illustrates the general
features. The emergence of a narrow peak and its rapid shift
toward high velocities for $x>x_{2}$ are clearly seen. The horizontal lines
indicate the points $x_{1}$-$x_{6}$ corresponding to the curves in
Fig.\ \ref {figure2}(a) and the arrows in the potential energy in Fig.\ 
\ref{figure1}(a).

The observed features are a signature of \emph{quasi-ballistic transport}.
The high-velocity peak in the distribution corresponds to \emph{ballistic}
electrons which, initially injected over the source-channel barrier, due to
the rapid and large (compared to $k_{B}T$)\bigskip\ potential drop and
corresponding strong inhomogeneous electric field, are accelerated to high
(positive) velocities. As discussed before, in the process, the mean free
path of  backscattered electrons is strongly reduced by the large
electric field, and thus, the ``ballistic'' peak reaches very high
velocities, and penetrates deep into the channel region,
as seen in Figs.\ \ref{figure2}(a) and \ref{figure2}(b). However, in
addition to 
the narrow ``ballistic'' peak, a broad distribution of electrons occupy the
low and negative velocity part of the phase space. This component of the
distribution function becomes more and more pronounced beyond the channel
mid-point. The origin of this broadening at the drain end of the
channel is due to electrons with
$v<0$, injected
from a near-equilibrium distribution in the drain contact which penetrate the
channel and contribute to the negative and low velocity components of the
distribution function. Within the channel, the inhomogeneous electric field
causes a non-uniform dependence of the mean free paths of the $n^{+}$ and $%
n^{-}$ components of the electron density, as discussed previously, which in
turn creates a feedback to the electrostatics due to the charge
(re)distribution. It is thus evident from Figs.\ \ref{figure2}(a) and \ref%
{figure2}(b) that the microscopic details of the influence of scattering in a
submicron channel can be obtained only from the full non-equilibrium
distribution function, which can only be obtained by the self-consistent
solution of the BTE.

\subsection{Crossover between quasi-ballistic and diffusive transport}
\label{crossover}
In the following, we will discuss the crossover between the quasi-ballistic
and diffusive regimes of transport in terms of numerical calculations and a
study of the microscopic signatures discussed above. We will focus on
the quantities discussed in the previous section and investigate their
dependence on an externally applied bias voltage, as well as the
dependence on the channel length. We
have performed calculations for different structures with channel lengths
ranging between 0.2 to 4 $\mu $m. In order to exclude the influence of
finite contact size on the results, we have performed and compared
studies using two definitions of the channel length variations. In one
case, which we simply call the constant contact (CC) size case, we
keep the size of the contacts fixed, while varying the channel length,
thus effectively changing the total sample size. The second case,
which we simply call the constant sample (CS) size case, we assume
that the total sample size is constant, such that the variation of the
channel length effectively reduces the contact size. In both cases, as
will be seen in the presented results, the contact sizes are found to
be large enough not to influence the main features of our study.

\subsubsection{Electrostatics}

\begin{figure}[t]
\par
\begin{center}
\scalebox{0.4}{\epsfig{file=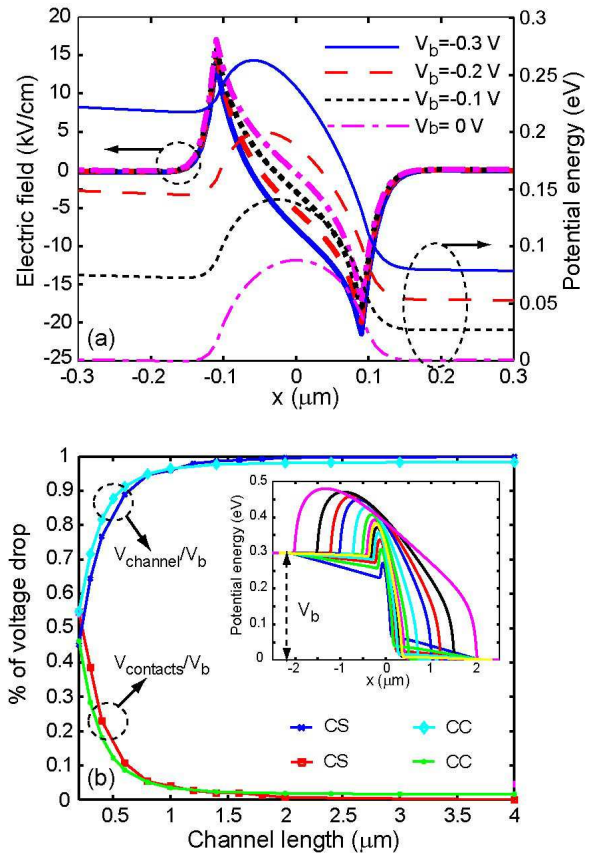}}
\end{center}
\caption{(Color online) (a) Potential energy (thin lines) and electric
  field (thick
  lines) profiles for different external bias voltages, for the
  same parameters and structure studied in Figs.\ \ref{figure1} and
  \ref{figure2}. (b) Fraction of the potential drop that occurs across
  the contacts, $V_{contacts}/V_{b}$, and the channel,
  $V_{channel}/V_{b}$, as a function of channel length. Inset shows
  the spatial dependence of the potential energy profile for channel
  lengths in the range $L_{c}=0.2-4$ $\mu$m. The different curves
  correspond to the constant contact (CC) and constant sample (CS)
  size approaches to the channel length variation in the
  calculations.}
\label{figure3}
\end{figure}

In Figs.\ \ref{figure3}, we show our calculated results of the
electrostatics for different applied bias voltages and different
channel lengths. In Fig.\ \ref{figure3}(a), the potential energy
profiles (thin lines) and electric field
distributions (thick lines) are shown for an $N^{+}/N/N^{+}$ structure
with dimensions 2.9/0.2/2.9 $\mu $m and the same parameters used in
the previous 
results, calculated at the applied bias voltages
$V_{b}=0,-0.1,-0.2,-0.3$ V. Even in the absence of an externally
applied field, a large
potential barrier and a corresponding strong and inhomogeneous electric
field builds up in the channel region due to the interdiffusion of electrons
from the highly doped contact regions to the lightly doped channel region.
As an external bias is applied to the sample boundaries at $\pm 3$ $\mu $m,
the majority of the potential drops over the channel region. We note that,
for the considered bias range, $V_{b}=-0.1$ to $-0.3$ V, the fraction of the
potential drop that takes place over the channel region is effectively
the same and $\approx 50$\%. The electric field, although
varying withing the same range of $\approx -25$ to $%
\approx $15 kV/cm has an increasingly stronger spatial dependence with
increasing bias voltage.

To investigate the crossover between the diffusive and quasi-ballistic
regimes of transport in the electrostatics, we have studied the
electrostatic response for structures with different
channel lengths, ranging between $L_{c}=0.2$ and 4 $\mu$m. The
potential energy 
profiles for different channel lengths (calculated for the CS case) at
$V_{b}=-0.3$ V
are shown in the inset of Fig.\ \ref{figure3}(b). It is evident from
the results that, although
most of the potential drop, and hence resistance, occurs over the
channel region, the fraction of the potential drop associated with the
channel region strongly depends on the channel length.

In Fig.\ \ref{figure3}(b) we show the fraction of the potential drop that
occurs across the channel, $V_{channel}/V_{b}$, and contact,
$V_{contacts}/V_{b}$, regions, respectively, for
different channel lengths. For long channels, virtually all of the
applied voltage drops across the channel region, irrespectively of the
magnitude of the applied voltage (comparison of the results calculated
for bias voltages in the range $V_{b}=-0.3$ to $V_{b}=-0.1$ V show
that there are very small variations in the length dependence and
magnitude of the voltage distribution in the system, within the
considered range of bias voltage). However, as the channel length is
successively decreased below $\approx 1.5$ $\mu$m, the fraction
$V_{channel}/V_{b}$ gradually
decreases, while $V_{contacts}/V_{b}$ gradually increases, to the point
where, for $L_{c}<0.3$
$\mu$m, the majority of the potential drop occurs across the contact
regions. Note that the calculated curves for the CC and CS size cases
are almost identical, showing that no contact size effects are
responsible for the observed features.

The observed results and dependence of $V_{channel}/V_{b}$
and $V_{contacts}/V_{b}$ show that a crossover occurs around $L_{c}\approx
1.5$ $\mu$m and that for shorter channel lengths, ballistic effects
become important. With decreasing channel length, the transfer of the
voltage drop from the channel region to the contact regions, indicates
that the transport becomes increasingly ballistic and it is expected
that for very short channels, the majority of the voltage drop will
occur across the contact regions, and only a very small (but finite)
resistance will occur within the channel region. The observed features
in the electrostatic response are a {\em signature of the crossover
  between the diffusive and quasi-ballistic regimes of transport}. 

\subsubsection{Electron density}

\begin{figure}[t]
\par
\begin{center}
\scalebox{0.4}{\epsfig{file=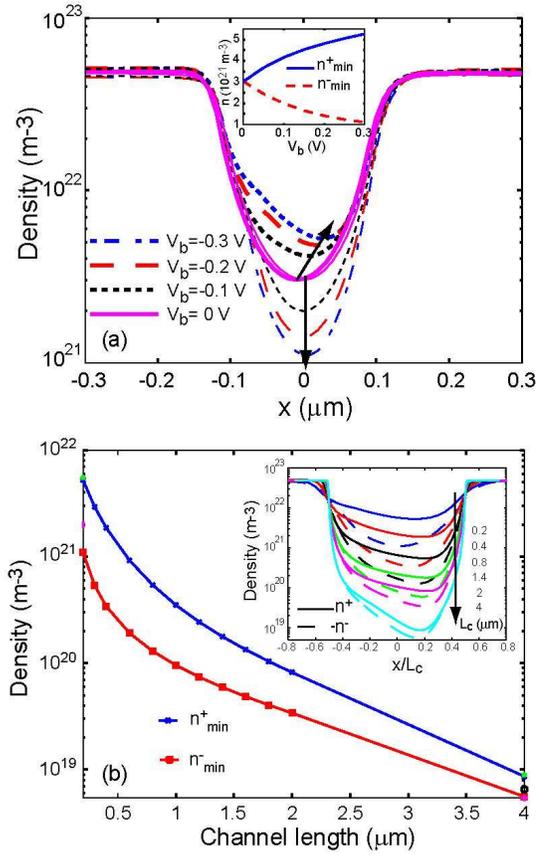}}
\end{center}
\caption{(Color online) (a) Electron densities, $n^{+}$ (thick lines) and $n^{-}$
  (thin lines) for the same structure and bias voltages as in
  Fig.\ \ref{figure3}. Inset shows the minimum densities, $n^{+}_{\min}$
  (solid line) and $n^{-}_{\min}$ (dashed line) as a function of
  $V_{b}$. (b) Minimum channel densities, $n^{+}_{min}$ and
  $n^{-}_{min}$, as a function of channel length. Inset shows electron
  density profiles, $n^{+}$ (solid lines) and $n^{-}$ (dashed lines),
  as a function of $x/L_{c}$, for different channel lengths, $L_{c}$.}
  \label{figure4}
\end{figure}

Next, we perform a similar study of the electron densities $n^{+}$ and
$n^{-}$, defined earlier in the text. In Fig.\ \ref{figure4}(a) we show
the electron density around the
channel region for the same structure as in Fig.\ \ref{figure3}(a),
calculated at the same bias voltages. The
thick and thin lines correspond
to the $n^{+}$ and $n^{-}$ components of the total density,
respectively. At $V_{b}=0$ the
two components are naturally centered, although slightly shifted from
each other, around the 
channel mid-point. The slight shift is due to the fact that, as seen in Fig.\
\ref{figure3}(a), the electric field is non-zero everywhere but at
mid-channel, and
anti-symmetric with respect to the channel center. Hence, the electron
distributions in each half of the channel consist of a slightly shifted MB
distribution. In the channel half close to the source, the distribution is
shifted toward negative velocities, whereas the distribution in the other
half is shifted toward positive velocity, see Fig.\ \ref{figure6}(d) for a
contour plot of the electron distribution around the channel region at
zero bias.
Consequently, $n^{+}-n^{-}\neq 0$ around the channel region (except
at the channel mid-point), where the
built-in field is non-zero. Note, however, that the {\em total}
electron density is entirely symmetric along the $x$-direction (not
shown here), as required since $V_{b}$=0.

As a bias voltage is applied, two main observations can be made. First, the
difference between the two components, $n^{+}-n^{-}$, increases considerably
with increasing $\left\vert V_{b}\right\vert $, which is also shown in the
inset to the figure for the minimum densities, $n_{\min }^{+}$ and $n_{\min
}^{-}$, along the channel. Second, the spatial dependence of $n^{+}$ is very
different from that of $n^{-}$, and more importantly varies considerably for
different applied voltages. In particular, the position of $n_{\min }^{+}$
changes significantly with $V_{b}$, whereas the position of $n_{\min }^{-}$
is roughly bias independent, as indicated by the schematic lines in the
figure. This is an indication that the efficiency of the scattering is
different for the $v^{+}$ and $v^{-}$ electrons. Interestingly, the
relative excess of $v^{+}$ electrons in the channel region reaches its
maximum around mid-channel (not shown), as previously seen in
Fig.\ \ref{figure1}(b) for $V_{b}=-0.3$ V, and the position is relatively
insensitive to the applied bias voltage. This is another indication
that the behavior in the channel is controlled by the intrinsic fields
and doping profiles rather than the externally applied electric fields.

As the channel length is increased (for $V_{b}=-0.3$ V) the spatial
dependence of the electron densities $n^{+}$ and $n^{-}$ changes
significantly. In the inset of Fig.\ \ref{figure4}(b) we show $n^{+}$
(solid lines) and $n^{-}$ (dashed lines) for different channel lengths
ranging between $L_{c}=0.2-4$ $\mu$m, as a function of $x/L_{c}$. In
addition, Fig.\ \ref{figure4}(b) shows the dependence of the minimum
densities, $n^{+}_{\min}$ and $n^{-}_{\min}$ as a function of the
channel length, calculated at $-0.3$ V for the CS case (we verified
again that the contact size and length scaling scheme do not influence
the results qualitatively).

It is seen that: First, as the channel length increases, the electron
density in the channel region decreases, approaching the value of the
background doping, $10^{19}$ m$^{-3}$. Second, the large magnitude
difference between $n^{+}_{\min}$ and $n^{-}_{\min}$, observed for short
channel lengths, decreases. Third, a crossover from a superexponential
to an exponential dependence of $n^{+(-)}_{min}$ on the channel length
is observed around $\approx 1.5$ $\mu$m. The latter observation we
again interpret as a signature of a transition between the quasi-ballistic and
diffusive regimes of transport, and occurs around similar channel
lengths as observed in the length dependence of the electrostatic
response, discussed in Fig.\ \ref{figure3}(b). For long channel
lengths, the density dependence approaches the purely diffusive regime
in which the $n^{+}$ and $n^{-}$ have a similar spatial dependence and
approach the values of the background doping. 

\begin{figure}[t]
\par
\begin{center}
\scalebox{0.4}{\epsfig{file=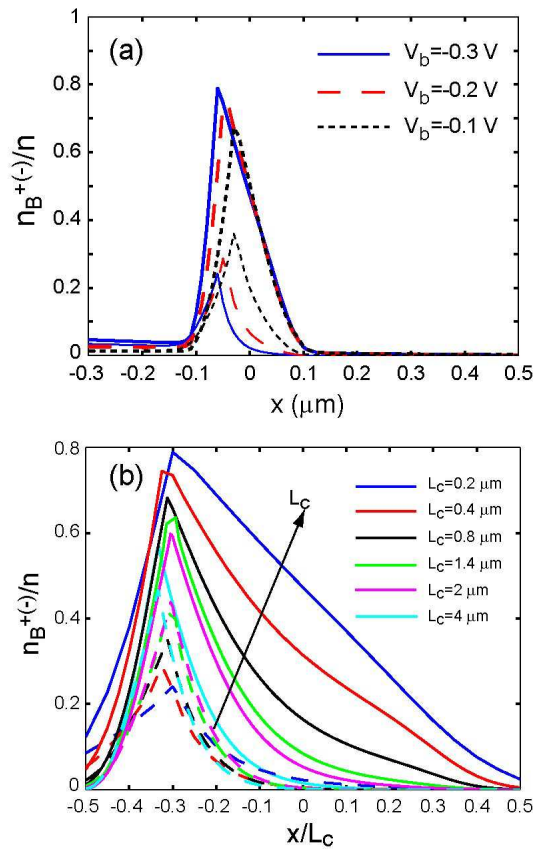}}
\end{center}
\caption{(Color online) ``Ballistic'' electron densities, $n^{+}_{B}$
  and $n^{-}_{B}$
  as defined in the text, for (a) the same structure
  and applied voltages as in Fig.\ \ref{figure4}(a), and (b) different
  channel lengths, $L_{c}$, vs. relative channel position, $x/L_{c}$.}
  \label{figure5}
\end{figure}

It is interesting to compare the results for the total electron
densities, obtained by the integration of the electron distribution over
the entire velocity space, see eq. (\ref{totaldens}), with the
``ballistic'' electron densities,
defined in eq. (\ref{ballistic density}). In Fig.\ \ref{figure5} we show the
fraction of positive and negative velocity components, $n^{+}_{B}$ and
$n^{-}_{B}$, of $n_{B}$ defined in eq. (\ref{ballistic density}),
relative to the total density.

In Fig.\ \ref{figure5}(a) we show the results for the same 200 nm
channel length structure studied in Figs.\ \ref{figure1},
\ref{figure2}, \ref{figure3}(a), and \ref{figure4}(a), calculated for
the bias voltages $V_{b}=-0.1$ $-0.2$ and $-0.3$ V. With the decrease
of bias voltage, the two peaks corresponding to $n^{+}_{B}/n$ (thick
lines) and $n^{-}_{B}/n$ (thin lines) are shifted toward the middle of
the channel, and decrease and increase, respectively, as expected by
the smaller asymmetry and magnitude of the electric field. We note that,
throughout the presented bias voltage range, $n^{+}_{B}/n$ decays
approximately linearly on both sides of the peak centered at the top of
the potential barrier at the source-drain interface. The fraction of
negative ``ballistic'' electrons, $n_{B}^{-}/n$, on the other hand,
has a faster, exponential-like decay on the drain side of the potential
barrier, indicating diffusive-like transport.

The length dependence of the ballistic fraction of electrons is
found to show a distinct signature of the crossover between the
diffusive and quasi-ballistic regimes of transport, at similar channel
lengths where the crossover was observed in the electrostatic
response as well as the total electron density distribution. 

In Fig.\ \ref{figure5}(b), we show $n^{+}_{B}/n$ (solid lines) and
$n^{-}_{B}/n$ (dashed lines) as a function of $x/L_{c}$. Interestingly,
$n^{-}_{B}/n$ has similar decay lengths, relative to $L_{c}$, and an
almost vanishing contribution around mid-channel (with the exception
of the structure with $L_{c}=0.2$ $\mu$m). We attribute this behavior
to the diffusive nature of the transport of electrons with $v<0$,
since the resistance in the diffusive regime is linear with respect to
the channel length.

The positive
velocity ballistic electron fraction $n^{+}_{B}/n$ on the other hand
has: {\em i)} a similar decay length and length dependence with
$n^{-}_{B}/n$ for $L_{c}=4$ $\mu$m, {\em ii)} an increasing relative
decay length for decreasing channel length, yielding a significant
contribution throughout the full extension of the channel. More
importantly, it is interesting to note the transition between an
exponential-like dependence for long channel lengths, to an almost
linear dependence at shorter channel lengths, as a function of the
relative position, $x/L_{c}$, within the channel region. We interpret
this as yet another signature of the crossover between the diffusive
and quasi-ballistic regimes of transport.

\subsubsection{Electron distribution function}

\begin{figure}[t]
\par
\begin{center}
\scalebox{0.4}{\epsfig{file=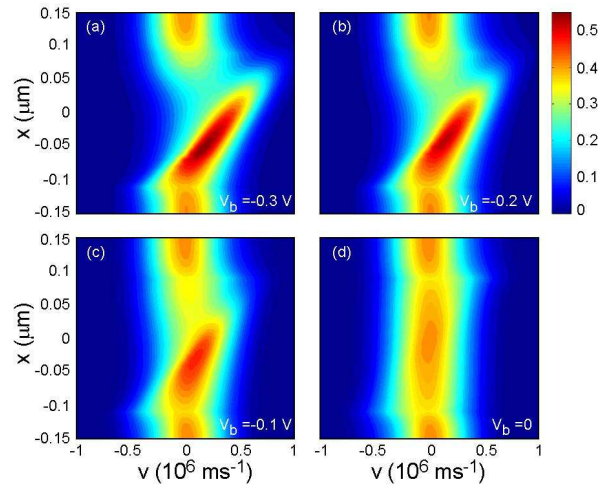}}
\end{center}
\caption{(Color online) Normalized electron distribution,
  $f(x,v)v_{th}/n(x)$, for 
  the structure with $L_{c}=200$ nm and the same parameters as studied in
  Figs.\ \ref{figure1} and \ref{figure2}, calculated at the bias
  voltages, (a) $V_{b}=-0.3$ V, (b) $V_{b}=-0.2$ V, (c) $V_{b}=-0.1$
  V, and (d) $V=0$.} \label{figure6}
\end{figure}

Having discussed the signatures of the quasi-ballistic and diffusive
regimes of transport in the electrostatic response and the electron density
distribution we now direct our attention to the full nonequilibrium
distribution function. In Figs.\ \ref{figure6} we
show the calculated normalized distribution function,
$f(x,v)v_{th}/n(x)$, for 200 nm channel
structures for different applied voltages. Figure \ref{figure6}(a)
shows the calculated normalized distribution at $V_{b}=-0.3$ and is
identical to the one shown in Fig.\ \ref{figure2}(b), and thus
displays the previously discussed ballistic, high-velocity peak
throughout the channel region. This peak shifts rapidly toward higher
velocities along the direction of transport and dominates the
distribution function well into the second half of the channel, close
to the drain side of the system. In the vicinity of the drain,
however, as previously discussed, drain-injected electrons spill over
into the channel and dominate the electron distribution. Note however
that the two contributions are well separated in phase-space and
thus can be easily identified.

As the bias voltage is decreased, the ballistic signatures become less
pronounced. Although there still are ballistic peak-like features for
$V_{b}=-0.2$ and $-0.1$ V, see Figs.\ \ref{figure6}(b) and (c),
respectively, the maximum velocity and the penetration depth of the
peak into the channel, i.e., the region in which it dominates the
electron distribution, are decreased. Furthermore, a smoother
transition occurs between the main peaks in the distribution on the
source and drain side of the channel [see Fig.\ \ref{figure6}(c)]
indicating an increasingly diffusive character of the transport. In
Fig.\ \ref{figure6}(d), for $V_{b}=0$ V, the distribution is as
expected symmetric along the direction of transport, but nevertheless
reflects the inhomogeneous properties of the sample, as seen in the
slightly shifted distributions around each of the two interfaces.

\begin{figure}[t]
\par
\begin{center}
\scalebox{0.4}{\epsfig{file=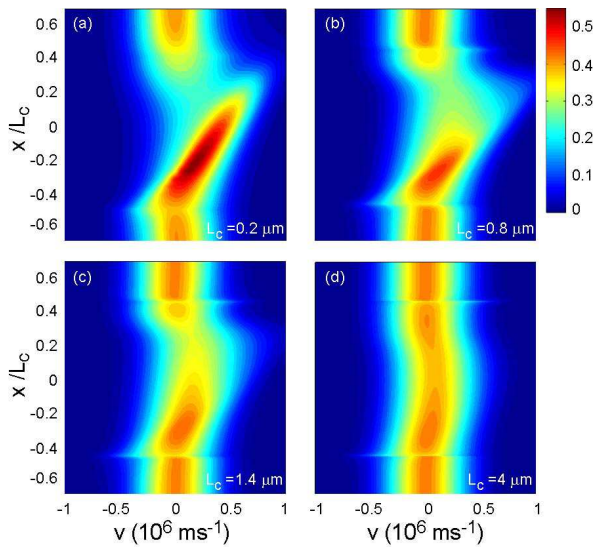}}
\end{center}
\caption{(Color online) Normalized electron distribution,
  $f(x,v)v_{th}/n(x)$, for 
  a structure with externally applied bias voltage $V_{b}=-0.3$ V,
  calculated for the channel lengths (a) $L_{c}=0.2$ $\mu$m, (b)
  $L_{c}=0.8$ $\mu$m, (c) $L_{c}=1.4$ $\mu$m, and (d) $L_{c}=4$ $\mu$m.}
  \label{figure7}
\end{figure}

In Fig.\ \ref{figure7} we show the normalized electron distribution
functions, calculated at $V_{b}=-0.3$ V, and for the channel lengths
$L_{c}=0.2$, 0.8, 1.4, and 4 $\mu$m. We again include the results for
the 200 nm long channel and $V_{b}=-0.3$ V biased system for
comparison in Fig.\ \ref{figure7}(a) [same as shown in
Fig.\ \ref{figure2}(b) and \ref{figure6}(a)]. Also, for the sake of
comparison between the different structures, we
show the normalized electron distribution function as a function of
velocity, $v$, and normalized spatial position in the channel,
$x/L_{c}$.

For $L_{c}=0.8$ $\mu$m, it is clear that the
quasi-ballistic features are persistent, although weaker, i.e., lower maximum
velocity and extension into the channel region (relative to the
channel length), as well as a smaller separation between the drain-
and source-injected electron distributions. Note again
that we plot the results against $x/L_{c}$ and that here, the
ballistic peak extends over 300 nm into the channel.

For $L_{c}=1.4$ $\mu$m, the distribution function shows significantly
reduced ballistic features. At this channel length, there is no clear
formation of a ballistic peak as in the previous
cases. Furthermore, the distribution has comparable width, only one
dominant peak throughout the entire channel region and a smooth
connection between the near-equilibrium source and drain
distributions. This indicates that the transport is
close-to-diffusive. For longer channels, the electron
distribution becomes even more featureless as the transport
becomes more and more diffusive. In Fig.\ \ref{figure7}(d), the
distribution function for a $L_{c}=4$ $\mu$m channel based system is
shown for comparison. No ballistic features are present, the
velocities are low in comparison to the previous cases and the
distribution in the channel region, although broadened and shifted due
to the increased fields, is close to the electron distributions in the
source and drain contacts. For channel lengths shorter than
$L_{c}$=1.4 $\mu$m we have, however, seen a transition toward an increasingly
ballistic regime of transport, with features resembling the ones shown
in Figs.\ \ref{figure7}(a) and (b).

\section{Summary and conclusions}
\label{summary}

In summary, we have performed a detailed microscopic study of
quasi-ballistic transport in deep submicron semiconductor
channels. In particular, we studied the crossover between the
quasi-ballistic and diffusive regimes of transport and identified
signatures in the electrostatic response, electron density and
full, nonequilibrium electron distribution function, that are due
to ballistic transport. The study was performed by numerically
solving the coupled, nonequilibrium Boltzmann-Poisson system of
equations, using a self-consistent direct method developed by us,
that enables us to obtain noise-free electron distributions
throughout the interesting submicron channel regions.

Semiconductor systems on the deep submicron and nanometer scale can in
general very easily be brought out of equilibrium due to their
reduced dimensionality. As shown in our model structures, even for
moderately downscaled channels and realistic doping levels, the
built-in fields are high and have a very strong spatial
dependence. Electron transport subject to these fields is
therefore highly out-of-equilibrium and drift-diffusion approaches
break down.

Our self-consistent calculations of the electron distribution
function in submicron GaAs channels using the Boltzmann-Poisson
equations show that the electron distribution is radically
different from a shifted Maxwell-Boltzmann distribution. The
distribution is far from Gaussian-shaped, has a large broadening,
and displays pronounced features, peaks and shoulders in the
high-velocity tail of the distribution. The high-velocity features
are signatures of ballistic transport, which occur due to the
large and strongly spatially varying electric fields. The large
broadening and significant contribution to the negative-velocity
component of the distribution are, however, in part due to
backscattered electrons in the channel, and partly due to
electrons injected from the drain.

We performed an analysis of the spatial density distributions for
electrons with positive and negative velocities, respectively. It was
shown that the relative scattering efficiency varies within the 
channel region and gives rise to asymmetric density distributions
for the forward and backward traveling electrons. We also defined and
studied a ``ballistic'' electron density, corresponding to electrons
with total energies higher than the energy of the top of the barrier
at the source-channel interface. It was shown that the spatial
dependence for the positive and negative velocity contributions to the
total ``ballistic'' density is radically different. This is due to the
large difference in the mean free path between the forward and
backward traveling electrons caused by the strong built-in electric
field. It was shown that a significant number of the source-injected
``ballistic'' electrons reach the channel-drain interface and
propagate into the drain. The relative fraction of ballistic electrons
in the drain is, however, very low due to the orders of magnitude
higher electron density in the drain contact. 

We have shown how these signatures are affected by an externally
applied bias voltage, and more importantly by the length of the
channel region. The effects of scattering have thereby explicitly
been illustrated.

In general, we have found that a crossover between the diffusive and
quasi-ballistic regimes of transport for the GaAs structures with the given
doping concentrations studied at room temperatures occurs around
channel lengths $L_{c}\approx 1.5$ $\mu$m, which is far longer than
the mean free path of $\approx 65$ nm determined by the thermal
velocity, $v_{th}$. The main parameters that influence this crossover
are the effective mass, the scattering time and the details of the
doping profile. Hence, a full self-consistent treatment such as ours
is needed to accurately model the crossover between the two regimes of
transport. 

In the
electrostatic response we found that the crossover manifests as a
redistribution of the voltage drop between the contacts and the
channel region. For large channel lengths, the transport is diffusive
and the majority of the potential drop occurs across the channel
region. For channel lengths smaller than $\approx 1.5$ $\mu$m a
monotonic redistribution of the potential drop from the channel region
to the contacts occurs, and a crossover between the two fractions 
occurs around $\approx 0.3$ $\mu$m, beyond which, for decreasing
channel lengths, the majority of the potential drop occurs across the
contact regions.

Similar crossovers were observed in the total and ``ballistic''
electron density distributions, $n^{(+,-)}$ and $n_{B}^{(+,-)}$, as
well as in the full, nonequilibrium distribution, $f(x,v)$. In
particular, we note that that the spatial distribution of $n_{B}^{+}$
relative to $x/L_{c}$ has an increasingly larger decay length and more
importantly shows a crossover between an exponential-like to a linear
decay rate around channel lengths at which the crossover between the
diffusive and quasi-ballistic regimes of transport were observed in
the electrostatic response. The spatial distribution of $n_{B}^{-}$,
however, has a relatively similar spatial distribution for different
channel lengths, relative to $L_{c}$, and displays exponential-like
decay indicative of diffusive transport. On a microscopic level, we
also showed 
explicitly how the signatures of the crossover manifest in the
nonequilibrium electron distribution, $f(x,v)$.
 
In general, scattering and quasi-ballistic transport in ultrasmall
semiconductor channels exhibit complex features due to the inherent nonlocal
nature of the problem. Our study has provided insight into the
microscopic signatures of the problem as well as of the crossover
between the diffusive and quasi-ballistic regimes of transport. We
note that some of the features reported in this paper resemble
experimental observations that may be explained within the presented
physical picture.

\acknowledgments
This work was supported by the Indiana 21st Century Research and
Technology Fund. Numerical calculations were performed using the
facilities at the Center for Computational Nanoscience at Ball State
University. Furthermore, the authors like to thank Dr. Savas Kaya
for fruitful discussions.


\begin{thebibliography}{99}
\bibitem{1DBTE} We assume that the electron distribution function can
  be decoupled into two equilibrium distribution functions in the $y$
  and $z$ directions of motion and one distribution function
  describing the electron motion in the $x$ direction where the
  electric field drives the electrons far-from-equilibrium.
\bibitem{scattering} $e^{-}-e^{-}$ scattering, while possibly comparable to 
polar-optical phonon scattering in the $n^{+}$ regions, is much weaker
in the channel  region due to the low doping concentration
($N_{D}=10^{13}$ cm$^{-3}$) in that region.
\end{thebibliography}
\end{document}